\documentclass[a4paper]{article}
\def\ud{{\rm{d}}}
\def\id{\,\ud}
\def\rr{{r^\ast}}

\begin{document}

\title{Diffusion of the scalar field energy due to the backscattering off
Schwarzschild geometry}

\author{Krzysztof Roszkowski\footnote{e-mail: {\tt roszkows@if.uj.edu.pl}}\\
{\it Institute of Physics, Jagellonian
University,} \\ {\it 30-059 Krak\'ow, Reymonta 4, Poland}}

\maketitle

{\abstract
This note tackles the problem of the backscattering of a mass-less scalar field
in the case
of Schwarz\-schil\-dean space-time. It shows that the effect depends both on
a distance from the horizon and on the wave length. The obtained estimates
significantly improve former results.
}

\section{Introduction}
The purpose of this note is to improve former estimates of the backscattering
effect \cite{SphericalScalarWavesAndGravity_MalcaNiallaChmaja} by adopting
a method used in \cite{DiffusionOfTheElectromagneticEnergy_Malca}.
The quantitative results obtained here
improve the former estimates by one order. They show that the backscattering
can be relevant only in case where the radiation source is placed close to
the horizon of a black hole or a very compact neutron star.
The effect depends on the wave frequency.

The result obtained analytically can be useful in order to test
numerical codes,
as suggested in \cite{SphericalScalarWavesAndGravity_MalcaNiallaChmaja}.
The present paper ignores the backreaction. We think, however,
that the method used here (or some of the estimates) can be useful
in the study of self-gravitating scalar fields.

\section{Formalism}
In following we use the Schwarzschildean line element:
\begin{equation}
\ud s^2=-\left(1-\frac{2m}{R}\right) \ud t^2 + \left(1-\frac{2m}{R}\right)^{-1}
\ud R^2 + R^2 (\ud \theta^2+ \sin^2 \theta \id \phi^2),
\end{equation}
where $t$ is a time coordinate and $R$ is a radial coordinate coinciding with
areal
radius; $0 \leq \phi < 2 \pi$ and $ 0 \leq \theta \leq \pi$.

Using a multi-pole expansion of scalar field in terms of scalar spherical
harmonics we obtain \cite{Wheeler1955}:
\begin{equation}
\left( -\partial_0^2 + \partial_\rr^2\right) \Psi_l =
\left(1-\frac{2m}{R} \right) \left[ \frac{2m}{R^3} + \frac{l(l+1)}{R^2} \right]
\Psi_l,
\label{MotionEquation}
\end{equation}
where $\rr \equiv R + 2m \ln(R/2m-1)$ is known as Regge-Wheeler
tortoise coordinate. The backreaction effect is neglected.

Define functions $\Psi_{ls}$ by recurrence relations:
\begin{eqnarray}
&
\partial_\rr \Psi_{l1} = - \frac{l(l+1)}{2}\Psi_{l0},
& \nonumber \\
&
\partial_\rr \Psi_{l(s+1)} = \frac{1}{2(s+1)} \left[ \left( s(s+1)-l(l+1)
\right) \Psi_{ls} - 2ms^2 \Psi_{l(s-1)} \right].
&
\end{eqnarray}
As one can easily see, in Minkowski space-time (which is a special case of
Schwarzschild space-time with $m=0$), the purely outgoing solution of equation
(\ref{MotionEquation}) is given by \cite{Bardeen1973}:
\begin{equation}
\widetilde{\Psi}_l (t,\rr) = \sum_{s=0}^l \frac{\Psi_{ls}(\rr-t)}{R^s}.
\label{MinkowskiSolution}
\end{equation}

Let us consider a function $\widetilde{\Psi}_l$ defined as in equation
(\ref{MinkowskiSolution}) and having compact support that is located in the
vacuum region outside radius $a>2m$. Assume that initial data of a solution
$\Psi_l$ of equation (\ref{MotionEquation}) are equal to $\widetilde{\Psi}_l$
and $\dot{\widetilde{\Psi}}_l$ at $t=0$.
It means that initially $\Psi_l$ is a purely outgoing wave.

In our calculations we use a following
\cite{DiffusionOfTheElectromagneticEnergy_Malca}
``decomposed'' solution $\Psi_l$:
\begin{equation}
\Psi_l(\rr,t)=\widetilde{\Psi}_l(\rr,t)+\delta_l(\rr,t),
\end{equation}
where $\delta_l$ is
unknown. Its presence is caused by the curvature of space-time.
From the preceding assumptions follows
$\delta_l(\rr,0)=\partial_0 \delta_l(\rr,0)=0$.

In the rest of the article we consider only a special case of the field
having no angular momentum, $l=0$. As consequence, we omit all subscripts
treating them as $0$.

\section{Energy estimate}
Let us define energy of the field $\Psi$, contained outside the ball
of radius $R$ as \cite{SphericalScalarWavesAndGravity_MalcaNiallaChmaja}:
\begin{equation}
E(R,t) = \int\limits_R^\infty \left[ \eta_r \left(\partial_r \Psi(r,t) -
\frac{\Psi(r,t)}{r} \right)^2 + \frac{\dot{\Psi}^2(r,t)}{\eta_r}
\right]\id r,
\end{equation}
where we used:
\begin{equation}
\eta_r = 1- \frac{2m}{r}.
\end{equation}
As one can easily check:
\begin{equation}
\left( \partial_0 + \partial_\rr \right) E(R,t) = -\left[
\dot{\Psi}(R,t) +\eta_R \left( \partial_r \Psi(R,t) -
\frac{\Psi(R,t)}{R} \right) \right]^2.
\end{equation}

Let $\Gamma_a$ be an outgoing null cone beginning at $a$. In Minkowski
space-time
all of the scalar field energy would be transported outside $\Gamma_a$.
In the Schwarzschild geometry some of wave will backscatter and enter
the region inside $\Gamma_a$. A fraction of the energy will be lost from
the main pulse. We will say that the energy ``diffuses'' inward through $\Gamma_a$.
We will estimate the amount of diffused energy.

In this
case, as on the geodesic $\Gamma_a$  $\widetilde{\Psi} =0$, we have
$\Psi = \delta$,
and therefore:
\begin{equation}
\left( \partial_0 + \partial_\rr \right)
E(R,t)|_{\Gamma_a} = 
-\left[\dot{\delta}(R,t) +\eta_R \left( \partial_r \delta(R,t) -
\frac{\delta(R,t)}{R} \right) \right]^2.
\end{equation}
If we introduce a function:
\begin{equation}
h_-(R,t) = \frac{1}{\eta_R} \left( \partial_0 + \partial_\rr \right)
\delta(R,t),
\end{equation}
then:
\begin{equation}
\left( \partial_0 + \partial_\rr \right)
E(R,t)|_{\Gamma_a} = 
\left[\eta_R h_-(R,t) - \eta_R \frac{\delta(R,t)}{R}\right]^2.
\end{equation}
The loss of the energy due to diffuse through $\Gamma_a$ reads:
\begin{equation}
\delta E = \int\limits_{(a,t=0)}^{(\infty,t=\infty)} \ud\rr
\left[\eta_R h_-(R,t) - \eta_R \frac{\delta(R,t)}{R}\right]^2.
\end{equation}

In order to find the estimate of this energy we will consider function, that
can be regarded as the ``energy of the $\delta$ field'':
\begin{equation}
H(R,t) = \int\limits_R^\infty \left[ \eta_r \left(\partial_r \delta(r,t) -
\frac{\delta(r,t)}{r} \right)^2 + \frac{\dot{\delta}^2(r,t)}{\eta_r}
\right]\id r.
\end{equation}
After some algebra we find that:
\begin{equation}
\left( \partial_0 + \partial_\rr \right) H(R,t) = -4m \int\limits_R^{\infty}
\dot\delta \frac{\widetilde{\Psi}(r,t)}{r^3} \id r - \left[ \eta_R h_-(R,t) -
\eta_R \frac{\delta(R,t)}{R} \right]^2,
\label{EqForH}
\end{equation}
what can be bounded from above by (we omit non-positive expressions):
\begin{equation}
\left( \partial_0 + \partial_\rr \right) H(R,t) \leq \left| 4m
\int\limits_R^{\infty} \dot\delta \frac{\widetilde{\Psi}(r,t)}{r^3} \id r
\right|.
\label{InequalityWithH}
\end{equation}
Notice that $\widetilde{\Psi}(r,t)=\widetilde{\Psi}(r_0,0)$,
where $(r,t) \in \Gamma_{r_0}$. Observe also that
$\frac{|\widetilde{\Psi}(r,t)|}{r} \leq \frac{|\widetilde{\Psi}(r_0,0)|}{r_0}$.
We can write:
\begin{equation}
\left|\frac{\widetilde{\Psi}(R_0,0)}{R_0} \right| = \left|
\frac{\widetilde{\Psi}(R_0,0)}{R_0} - \frac{\widetilde{\Psi}(b,0)}{b} \right|
=\left| \int\limits_{R_0}^{b} \partial_r \frac{\widetilde{\Psi}(r,t)}{r} \id r \right| .
\end{equation}
After the use of the Schwartz inequality, we have:
\begin{eqnarray}
&
\left|\frac{\widetilde{\Psi}(R,t)}{R}\right| \leq
\sqrt{\int\limits_a^b \frac{1}{r^2 \eta_r} \id r}
\sqrt{\int\limits_a^b \eta_r \left( \partial_r \widetilde{\Psi}(r,0) -
\frac{\widetilde{\Psi}(r,0)}{r}\right) \ud r} \leq
& \nonumber \\
&
\leq \sqrt{\frac{1}{2m} \ln \frac{\eta_b}{\eta_a}} \sqrt{E(a,0)}.
\label{BoundOnPsi}
&
\end{eqnarray}

Now we can use the Schwartz inequality in order to bound the right hand side of
(\ref{InequalityWithH}). We have:
\begin{equation}
\left( \partial_0 + \partial_\rr \right) H(R,t) \leq
4m \sqrt{\int_R^\infty \ud r \frac{\dot{\delta}^2 (r,t)}{\eta_r}}
\sqrt{\int_R^\infty \ud r \frac{\widetilde{\Psi}^2(r,t) \eta_r}{r^6}}.
\end{equation}
$\widetilde{\Psi}$ vanishes on a null cone $\Gamma_b$ directed outward
from $(b,0)$.
Using equation (\ref{BoundOnPsi}) and omitting non-positive terms in $H(R,t)$
we obtain:
\begin{equation}
\left( \partial_0 + \partial_\rr \right) H(R,t) \leq
4m \sqrt{H(R,t)} \sqrt{E(a,0)} 
\sqrt{\int_R^{R(b)} \frac{1}{2m} \frac{\eta_r}{r^4}
\ln \frac{\eta_b}{\eta_a}  \id r},
\end{equation}
where we changed the upper limit of the integral form $+\infty$ to $R(b)$,
which is the external end of the field's support.
We obtain:
\begin{eqnarray}
\left( \partial_0 + \partial_\rr \right) H(R,t) &
\leq &
4m \sqrt{H(R,t)} \sqrt{E(a,0)} \frac{1}{R^{3/2}}
\sqrt{\frac{1}{6m} \ln \frac{\eta_b}{\eta_a}} \times
\nonumber \\
&
&
\times \sqrt{1-\left(\frac{R}{R(b)}\right)^3 + 
\frac{6m}{4R(b)}\frac{R^3}{R(b)^3}-\frac{6m}{4R(b)}},
\end{eqnarray}
that can be bounded by (see \cite{DiffusionOfTheElectromagneticEnergy_Malca}
for proof that $\frac{R}{R(b)} \geq \frac{a}{b}$):
\begin{equation}
\left( \partial_0 + \partial_\rr \right) H(R,t) \leq
4m \sqrt{H(R,t)} \sqrt{E(a,0)} \frac{1}{R^{3/2}}
\sqrt{\frac{1}{6m} \ln \frac{\eta_b}{\eta_a}}
\sqrt{1-\left(\frac{a}{b}\right)^3},
\end{equation}

Applying $\partial_x \sqrt{f} = \frac{\partial_x f}{2\sqrt{f}}$ we finally
obtain:
\begin{equation}
\left( \partial_0 + \partial_\rr \right) H^{1/2}(R,t) \leq
2m \sqrt{E(a,0)} \frac{1}{R^{3/2}}
\sqrt{\frac{1}{6m} \ln \frac{\eta_b}{\eta_a}}
\sqrt{1-\left(\frac{a}{b}\right)^3}
.
\end{equation}
That gives us:
\begin{eqnarray}
H^{1/2}(R,t) & \leq &
2m \sqrt{E(a,0)} \sqrt{\frac{1}{6m} \ln \frac{\eta_b}{\eta_a}}
\sqrt{1-\left(\frac{a}{b}\right)^3}
\int\limits_a^R \ud\rr \frac{1}{r^{3/2}} \leq
\nonumber \\
&
\leq &
2m \sqrt{E(a,0)} \sqrt{\frac{1}{6m} \ln \frac{\eta_b}{\eta_a}}
\sqrt{1-\left(\frac{a}{b}\right)^3}
 \times
\nonumber \\
&
&
\times \sqrt{\frac{2}{m}}\left[ \tanh^{-1} \sqrt{\frac{a}{2m}} -
\tanh^{-1} \sqrt{\frac{R}{2m}} \right].
\end{eqnarray}

Let's come back to equation (\ref{EqForH}):
\begin{equation}
\left( \partial_0 + \partial_\rr \right) H(R,t) = -4m \int\limits_R^{\infty}
\dot\delta \frac{\widetilde{\Psi}(r,t)}{r^3} \id r - \left[ \eta_R h_-(R,t) -
\eta_R \frac{\delta(R,t)}{R} \right]^2.
\end{equation}
After integrating both sides along null cone originating at $a$, we arrive at:
\begin{equation}
H(\infty,\infty)-H(a,0) \leq \int\limits_{(a,t=0)}^{(\infty,t=\infty)} \id\rr
\left[ -4m \int\limits_R^{\infty} \dot\delta \frac{\widetilde{\Psi}(r,t)}{r^3}
\id r \right] - \delta E.
\end{equation}
Because of the vanishing of both $\delta$ and all of its derivatives at $t=0$
we have $H(a,0) = 0$. $H$ is manifestly non-negative, hence:
\begin{equation}
\delta E \leq \int\limits_{(a,t=0)}^{(\infty,t=\infty)} \id\rr
\left[ -4m \int\limits_R^{\infty} \dot\delta \frac{\widetilde{\Psi}(r,t)}{r^3}
\id r \right]. 
\end{equation}
Having found a bound on $H$, we obtain:
\begin{equation}
\left|
-4m \int\limits_R^{\infty} \dot\delta \frac{\widetilde{\Psi}(r,t)}{r^3}
\id r \right|
 \leq
4m \sqrt{H(R,t)} \sqrt{E(a,0)} \frac{1}{R^{3/2}}
\sqrt{\frac{1}{6m} \ln \frac{\eta_b}{\eta_a}}
\sqrt{1-\left(\frac{a}{b}\right)^3}.
\end{equation}
Thus finally we arrive at:
\begin{eqnarray}
\delta E &
\leq
 &
m \left( 1 -  \left( \frac{a}{b}\right)^3 \right) \ln \frac{\eta_b}{\eta_a}
E(a,0)\frac{4}{3} \times
\nonumber \\
&
&
\times \int\limits_a^\infty \ud \rr  \frac{1}{r^{3/2}}
\sqrt{\frac{2}{m}}\left[ \tanh^{-1} \sqrt{\frac{a}{2m}} -
\tanh^{-1} \sqrt{\frac{R}{2m}} \right]
\leq
\nonumber \\
&
\leq &
\frac{4 E(a,0)}{3 \eta_a} \left[ 1 -  \left( \frac{a}{b} \right)^3 \right]
\ln \frac{\eta_b}{\eta_a} \ln \frac{1}{\eta_a},
\end{eqnarray}
where, during the integration, we put $\frac{1}{\eta_a}$ instead of
factor $\frac{1}{\eta_r}$ that appears when one changes the integration
variable $\rr$
to $r$. Although this operation makes our boundary a little
bit worse, it enables to write the final result in a much more compact way.

\section{Comments}
In this note we arrived at a very simple expression bounding energy lost
by an outgoing mass-less scalar field. It is worth to say few words about
some consequences of the result.

There is a connection between wave frequencies of the pulse
and its width, i.e., the more narrow the pulse is the shorter the wave length.
The width of the pulse depends on the relation between $a$ and $b$, when the
pulse is located in a very narrow region $a \to b$. This means
$(1-\frac{a^3}{b^3})\to 0$,
and we find that short waves are scattered much less than long ones.

If we are looking into a region placed far away from a horizon, where
$a \gg 2m$, we can even simplify found expression, expanding functions in the
last bracket. We obtain:
\begin{equation}
\delta E \leq \frac{4}{3} E(a,0) \left( \frac{2m}{a} \right)^2
\left( 1 - \frac{a}{b} \right) \left[ 1-\left( \frac{a}{b} \right)^3 \right].
\end{equation}

\section*{Acknowledgments}
The author is grateful to Edward Malec for
suggesting the topic, many discussions and help in preparation of
the manuscript.

\end{document}